# Do single photons tunnel faster than light?

Herbert G. Winful

EECS Department, University of Michigan, 1301 Beal Ave., Ann Arbor, MI 48109

## ABSTRACT

Experiments done with single photons in the early 1990's produced a surprising result: that single photons pass through a photonic tunnel barrier with a group velocity faster than the vacuum speed of light. This result has stimulated intense discussions related to causality, the speed of information transfer, the nature of barrier tunneling and the meaning of group velocity. The superluminality of tunneling photons is now textbook material, although the authors note that controversy still remains. Another paradoxical result, known as the Hartman effect, is that the tunneling time of the photons becomes independent of barrier length in the limit of opaque barriers. In this paper we examine the meaning of group velocity in the context of barrier tunneling. We ask whether a single tunneling photon can be described by a group velocity and whether the short group delays imply superluminal group velocity. We resolve the paradox of the Hartman effect and show that the predicted and measured group delays are not transit times but photon lifetimes.

## 1. INTRODUCTION

How long does it take a particle or wave packet to tunnel through a barrier? This question has occupied physicists for decades and yet there is still no definitive, universally accepted answer. In the early 1930's, MacColl found, on the basis of a wave packet solution of the time-dependent Schrödinger equation, that "tunneling takes no appreciable time" [1]. In other words, the peak of the transmitted wave packet departs the exit of the barrier at about the same time the peak of the incident packet arrives at the entrance. Hartman later found that there is finite delay, the group delay, given by the energy derivative of the steady state transmission phase shift [2]. This delay, however, saturates with barrier length, a phenomenon that has been termed the Hartman effect. If the group delay is taken as a transit time, the implication is that the group velocity calculated by dividing barrier length by transit time can exceed the speed of light. Indeed it would grow with barrier length, ultimately becoming unbounded. This paradoxical effect has been a mystery for years. In the early 1990's the classic single-photon experiments of Steinberg, Kwiat, and Chiao demonstrated apparent superluminal group velocities of about 1.7$c$ [3]. Later experiments with classical light pulses confirmed the Hartman effect [4]. As a result, it is now widely believed that tunneling photons travel with superluminal group velocity through photonic bandgaps.

Our recent work contradicts the notion that photons, classical electromagnetic waves, or matter waves tunnel with superluminal group velocity [5-14]. It shows that tunneling is a quasi-static phenomenon and cannot be understood with the "group velocity" concepts appropriate for pulse propagation studies. It shows that the group delay is not a transit time but the lifetime of stored energy (or stored probability density). Since it is not a transit time it cannot be used to assign a "group velocity" for tunneling. We have shown that the group delay is proportional to the time-average stored energy in the barrier. We have used the proportionality between the stored energy and the group delay to explain the riddle of the Hartman effect and to reinterpret the tunneling time experiments without appealing to superluminality or some hypothetical reshaping process that leaves the pulse unchanged.

## 2. TUNNELING IS A QUASI-STATIC PHENOMENON

It is important to note that true tunneling is a *quasi-static* phenomenon requiring pulses much longer than the length of the barrier [5,9,10]. Changes in the input pulse envelope occur on a time scale long compared to the time it takes information about the change, traveling at $c$, to reach the exit. As a result the transmitted pulse is accurately described by $A(L,t) = T_0 A(0, t - \tau_g)$, where $T_0$ is the steady state transmission at the center frequency of the wave packet. It is seen that every part of the delayed input pulse experiences the same transmission. This is why the transmitted pulse is not shortened or distorted or reshaped. The quasi-static nature of tunneling arises from the requirement that the spectrum of the input pulse be narrow compared to the width of the stopband. The group delay $\tau_g$ in tunneling is shorter than the transit time L/c of a light front and is a very small fraction of the pulse temporal width.

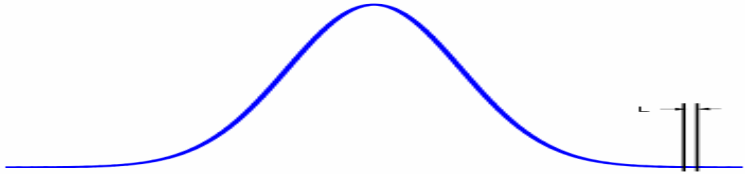

Fig. 1. Quasi-static nature of tunneling. The wave packet is much longer than the barrier so that at any instant, near steady-state conditions obtain.

## 3. THE GROUP DELAY IN TUNNELING IS *NOT* A TRANSIT TIME BUT A LIFETIME

It is universally assumed that the group delay in tunneling is a transit time of a wave packet from input to output. We have critically examined this assumption and have shown that the group delay is not a transit time but the lifetime of stored energy (or stored probability) escaping through both ends of the barrier [6]. It is in essence a cavity lifetime. Because it is a lifetime and not a transit time, it should not be linked to a velocity. The group delay is a property of an entire wave function with transmitted and reflected components. It cannot be assigned to just the forward propagating part. It has the same status as the dwell time which is also a lifetime and not a transit time.

## 4. THE ORIGIN OF THE HARTMAN EFFECT IS THE SATURATION OF STORED ENERGY

The Hartman effect is the saturation of the group delay with barrier length. If the group delay is interpreted as a transit time, its saturation implies that tunneling velocities can become superluminal and, indeed, unbounded as the barrier length is increased. Thus the wave packet must somehow know that the barrier length has increased and therefore it should speed up to get to the exit. This does not make sense. On the other hand, if the group delay is not a transit time but a lifetime, then its saturation is easy to explain. The group delay is proportional to the stored energy in the barrier. If the stored energy saturates then so must the group delay. Because of the exponential decay of the energy density with distance, beyond a 1/e distance, it does not matter how long the barrier is: all the energy is stored within the 1/e distance. Thus the Hartman effect is due to the saturation of stored energy with barrier length [7].

# 5. REINTERPRETATION OF TUNNELING TIME EXPERIMENTS

The experimental result Steinberg, Kwiat, and Chiao has been taken to mean that tunneling photons travel faster than light, with a group velocity of $1.7c$. The widespread belief in the physics community is that if we launch two identical photons and one passes through a barrier while the other passes through an equal length of free space, the photon that went through a barrier will arrive first at the detector. However our work offers an explanation of the SKC experiment that does not require superluminal velocities [5].

First, we note that the experimental conditions satisfy the quasi-static conditions. To explain the SKC results we consider the photons as simply modes of the electromagnetic field that are generally independent and hence do not exhibit second-order interference. They can nevertheless exhibit fourth-order interference which is monitored through a Hong-Ou-Mandel interferometer. The detection probabilities depend on the relative phase between the two photons. However, the phase accumulated by a mode in any region of space is proportional to the energy stored in that region:

$$\phi_0 = \frac{U}{P_{in}}\Omega,$$

where $U$ is the stored energy, $P_{in}$ is the input power, and $\Omega$ is the carrier frequency minus the Bragg frequency. The phase is linear in frequency, which means that in the time domain there is a pure delay without distortion. This is why both cw and pulsed measurements under quasi-static conditions yield the same result for the group delay. More importantly, for a given input power the phase is proportional to the stored energy. In the stop band of a photonic band gap structure the stored energy is reduced below its free space value (as a result of destructive interference) and this leads to a reduction in accumulated phase below what would have been gained through free propagation. In order to make up for the loss of phase, the path length external to the barrier has to be increased, thereby adding some propagation phase. The phase lag in the barrier is due to the finite response time of the structure acting as a lumped element. It is the lifetime of stored energy leaving both ends of the structure, most of it in the backward (reflection) direction. The phase lag in free space results from the need to transport stored energy out of the region in the forward direction before fresh energy can enter. It should be noted that the experiment simply measured the shift in a mirror position and not a velocity. This shift corresponds to the difference in stored energies and does not imply that anything was transported with superluminal velocity.

In summary, what is measured in tunneling time experiments is the time it takes for the energy stored in the barrier to leak out of both ends of the barrier. It is identical to the dwell time and should not be used to calculate a propagation velocity. The measured group delay is the photon lifetime within the stop band. This lifetime can be arbitrarily short for a highly reflective barrier. It does not imply that anything is traveling faster than light. My conclusion is that photons do not tunnel with superluminal group velocity.